\documentclass{article}

\usepackage{arxiv}

\usepackage[utf8]{inputenc} % allow utf-8 input
\usepackage[T1]{fontenc}    % use 8-bit T1 fonts
\usepackage{hyperref}       % hyperlinks
\usepackage{url}            % simple URL typesetting
\usepackage{booktabs}       % professional-quality tables
\usepackage{amsfonts}       % blackboard math symbols
\usepackage{nicefrac}       % compact symbols for 1/2, etc.
\usepackage{microtype}      % microtypography
\usepackage{lipsum}
\usepackage{graphicx}
\graphicspath{ {./images/} }
\usepackage{authblk}
\usepackage{caption}
\usepackage{subcaption}
\usepackage{multirow}
\usepackage[style=ieee,backend=biber]{biblatex}
\usepackage{hyperref}
\usepackage{color}
\hypersetup{%
colorlinks=true,
linkcolor=blue,
citecolor=blue,
urlcolor=blue}

\usepackage{hyperref}
\usepackage{bookmark}
\usepackage{etoolbox}
\makeatletter

\usepackage{float}
\restylefloat{table}

\def\tsc#1{\csdef{#1}{\textsc{\lowercase{#1}}\xspace}}
\tsc{WGM}
\tsc{QE}
\tsc{EP}
\tsc{PMS}
\tsc{BEC}
\tsc{DE}

\title{CT-xCOV: a CT-scan based Explainable Framework for COVid-19 diagnosis}

\author{Ismail ELBOUKNIFY\textsuperscript{1}\thanks{ielbouknify@gmail.com}, Afaf BOUHOUTE\textsuperscript{2}, Khalid FARDOUSSE\textsuperscript{2}, Ismail BERRADA\textsuperscript{3}, Abdelmajid BADRI\textsuperscript{1}}

\affil{\textsuperscript{1} EEA\&TI Laboratory, Hassan 2 University, Casablanca, Morocco\\
\textsuperscript{2} Sidi Mohamed Ben Abdellah University, Faculty of Science, Fez, Morocco\\
\textsuperscript{3} Mohammed VI Polytechnic University, School of Computer Science, Benguerir, Morocco}

\begin{document}
\maketitle
\begin{abstract}
In this work, CT-xCOV, an explainable framework for COVID-19 diagnosis using Deep Learning (DL) on CT-scans is developed. CT-xCOV adopts an end-to-end approach from lung segmentation to COVID-19 detection and explanations of the detection model's prediction.
For lung segmentation, we used the well-known U-Net model. For COVID-19 detection, we compared three different CNN architectures: a standard CNN, ResNet50, and DenseNet121. After the detection, visual and textual explanations are provided. For visual explanations, we applied three different XAI techniques, namely, Grad-Cam, Integrated Gradient (IG), and LIME. Textual explanations are added by computing the percentage of infection by lungs. To assess the performance of the used XAI techniques, we propose a ground-truth-based evaluation method, measuring the similarity between the visualization outputs and the ground-truth infections. The performed experiments show that the applied DL models achieved good results. The U-Net segmentation model achieved a high Dice coefficient (98\%). The performance of our proposed classification model (standard CNN) was validated using 5-fold cross-validation (acc of 98.40\% and f1-score 98.23\%). Lastly, the results of the comparison of XAI techniques show that Grad-Cam gives the best explanations compared to LIME and IG, by achieving a Dice coefficient of 55\%, on COVID-19 positive scans, compared to 29\% and 24\% obtained by IG and LIME respectively. 
The code and the dataset used in this paper are available in the GitHub repository \cite{project}.
\end{abstract}

\section{Introduction}

The COVID-19 outbreak has put the medical researchers under an unprecedented spotlight. With the rapid transmission of the virus, identifying effective tools for early diagnosis becomes crucial to prevent further spread. So far, The Reverse Transcription Polymerase Chain Reaction (RT-PCR) has been widely used for detecting COVID-19. However, the difficulties posed by high false negative rates, slow processing times, inconsistent testing protocols, and detection sensitivity in RT-PCR tests \cite{hu2020weakly}, present significant obstacles to achieving early and accurate COVID-19 detection. Besides, Chest CT-scans have emerged as a crucial tool for both screening and diagnosing lung infections associated with COVID-19 \cite{fang2020sensitivity}. Chest CT findings reveal that CT-scans of COVID-19-infected patients show some visual signs such as ground glass opacity, consolidation, and pleural effusion \cite{ kwee2020chest}. Therefore, using a CT-scan can enable early detection of COVID-19 infection in patients with initially a negative RT-PCR examination or who are asymptomatic, both before and following the onset of symptoms\cite{hu2020weakly}.

With the recent breakthroughs in Artificial Intelligence (AI), more attention has been paid to using AI approaches to assist in making more accurate and reproducible radiology assessments \cite{hosny2018artificial}. Several studies have shown the efficacy of AI for accelerating the diagnosis of many diseases.  
Recently, Deep Learning (DL) models have become increasingly popular for CT-scan imaging-based diagnosis. These models have achieved a remarkable performance, improving the radiologist's performance in detecting many diseases \cite{sun2021deep}.
However, this increased prediction accuracy is often associated with higher model complexity. 
The black-box nature of the DL models makes it difficult to comprehend their inner workings, which means that clinicians using DL-based diagnosis will be unable to understand the most important features that contribute to the diagnosis.
This issue increases skepticism around the clinical use of DL models and gives rise to a need for interpretability.

eXplainable Artificial Intelligence (XAI) is a new emerging domain, which brings together several processes and methods to provide human users with the ability to comprehend and trust the results of ML algorithms.
Recent research in medical imaging analysis is increasingly focused on creating human-understandable explanations of how a model makes specific individual predictions \cite{hakkoum2022interpretability}. These explanations are provided as visualization of the image regions that contribute the most to model predictions. While XAI methods have achieved a certain level of success, evaluating and quantifying their effectiveness is still a challenge.

The goal of this paper is to make use of the popular success of DL, the availability of CT images, and the increasing need for interpretability to build accurate and explainable COVID-19 detection systems. To this end,
we propose CT-xCOV, an end-to-end framework for explainable COVID-19 diagnosis. The framework adopts popular CNN architectures for lung segmentation and classification, and visual XAI methods for explanations. The CT-xCOV framework is validated using three popular COVID-19 datasets.
The key contributions of this paper can be outlined as follows:
\begin{itemize}
     \item %\textcolor{red}{Put something related to CT-xCOV framework end to end ...} 
     \textbf{CT-scan-based COVID-19 detection using Deep Learning (DL)}: we use and evaluate different DL architectures for lung segmentation and CT-scan classification. For segmentation, we trained the well-known U-Net model to generate lung masks. For classification, we evaluated different CNN architectures, including DenseNet and ResNet to classify the lungs and distinguish healthy and COVID-19 infected.
    % \textbf{Lung segmentation}: to prepare the CT images for the training process, segmentation is performed to identify the lung regions. For this task, the U-Net model has been used to generate the lung masks.
    %\item \textbf{Classification}: for COVID-19 detection, we evaluated different CNN architectures, including DenseNet and ResNet to classify the lungs and distinguish healthy and COVID-19 infected. 
    \item \textbf{XAI methods for an explainable detection}: To provide visual explanations about the COVID-19 predictions, we applied three different methods, namely Grad-CAM, Integrated Gradient, and LIME. A textual explanation about the degree of infection, based on this visualization, is also provided. %This explanation gives information about the degree of infection, computed as the size of the infected regions divided by the size of the lungs.
    
    \item \textbf{Explainability assessment}: to evaluate the explanations provided by the different methods, we propose a ground-truth-based evaluation method. This method compares the visualization output with the ground-truth infections. The performance is reported as an average of the similarity coefficient between visualization and ground-truth infection masks.
\end{itemize}

%structure
The remainder of this paper is structured into four sections.
In section \ref{meth}, we present an overview of the methodology of our approach and its different steps.
In section \ref{res}, the performed experiments are detailed, and the obtained results are discussed. Besides we report some related works on COVID-19 detection based on medical images in section \ref{relWork}.
Finally, section \ref{conc} draws conclusions and gives the perspectives.

\section{CT-xCOV : An end-to-end framework for eXplainable COVID-19 diagnosis}
\label{meth}

Following the recent advances in Trustworthy and Explainable Artificial Intelligence (XAI), in this section, we introduce CT-xCOV, our new framework for explainable and automated COVID-19 diagnosis. %The proposed CT-xCOV adopts deep learning models for COVID-19 diagnosis automation and XAI to provide trustworthy diagnoses. 
The section starts with an overview of the CT-xCOV architecture and follows with the research methodology adopted to address our goals.
%This section describes our research methodology to achieve an explainable diagnosis of COVID-19. First, we start with an overview of the proposed approach. Then, we introduce the datasets used and present the different steps, starting from data preprocessing and classification up to the predictions’ explanation and explanations assessment. 

\subsection{Architecture Overview}

CT-xCOV consists of four key components: 1) data preparation, 2) lung segmentation, 3) COVID-19 detection, and 4) the explainability component. The first component is responsible for collecting, preprocessing, and preparing the CT-scan dataset for the training process. As will be detailed later, to deal with the lack of a large COVID dataset, three different datasets were combined. The second component identifies the Region Of Interest (ROI), i.e. the lung in our case. The third component is in charge of classifying the lungs into two classes, Infected and Uninfected. We performed classification using different CNN architectures. Finally, the last component explains the model prediction and makes it more transparent. To this end, XAI methods are used to provide visual explanations about the model decision. Several XAI visual methods were compared and their performance was assessed. In the case of COVID-19 positive, textual explanation is also provided in forms of infection degree.

\begin{figure}[h]
    \centering
    \includegraphics[scale=0.7]{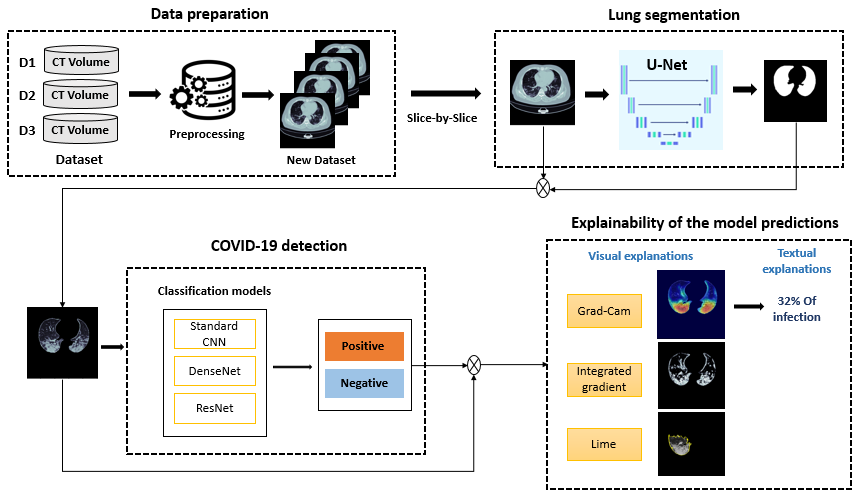}
    \caption{Overall Architecture of the proposed framework.}
    \label{fig:my_arch}
\end{figure}

Figure \ref{fig:my_arch} illustrates the general architecture and flow of the proposed framework. CT-xCOV is a CT-scan-based framework. A CT-scan volume contains a variable number of 2D axial slices (i.e. the number of slices in a CT scan may vary). Moreover, the shape, as well as the size of the lung, vary significantly between slices. Therefore, CT-xCOV adopts a 2D slice-based approach for COVID-19 detection. The slice-based approach takes as input individual CT slices and generates, for each slice, two outcomes:  the infection prediction, and the explanations of the reasons behind this prediction.
%Next, the different components are detailed in the next subsections.

\subsection{Data Preparation}
%\subsubsection{CT Datasets}
This first component of CT-xCOV deals with data preparation. Collected data is pre-processed and prepared considering different pre-processing techniques. 
As part of our framework design, we considered the following three public datasets:

%To deal with the lack of publicly available COVID-19 datasets, the  following three public datasets were used to train and evaluate the proposed framework: 
%We refer to the used datasets as $D_1$, $D_2$ and $D_3$. 
%The description of these datasets is given below:

\begin{itemize}%[noitemsep,nolistsep]
    \item \textit{\textbf{COVID-19 CT Lung and Infection Segmentation Dataset}} \cite{zenodo} (refereed to as $D_1$). It contains CT volumes of 20 patients, all infected with COVID-19. All the cases were labeled by two radiologists using three labels (left lung, right lung, and infections), and confirmed by a proficient radiologist.
    \item \textit{\textbf{COVID-19 CT segmentation dataset}} \cite{medicalsegmentation} (refereed to as $D_2$). It is a dataset of 100 axial CT images from over 40 COVID-19-positive patients. The lungs and infections were segmented by a radiologist.

    \item \textit{\textbf{MosMedData}} \cite{morozov2020mosmeddata} (refereed to as $D_3$). It contains CT-scans provided by hospitals in Moscow, Russia. It includes 1110 volumetric CT-scans of unaffected and affected cases. Annotations of infections for 50 case studies were performed by experts from the Moscow Health Care Department.

\end{itemize}
\begin{figure}[h]
    \centering
    %[width=0.6\textwidth]
    \includegraphics[width=0.9\textwidth]{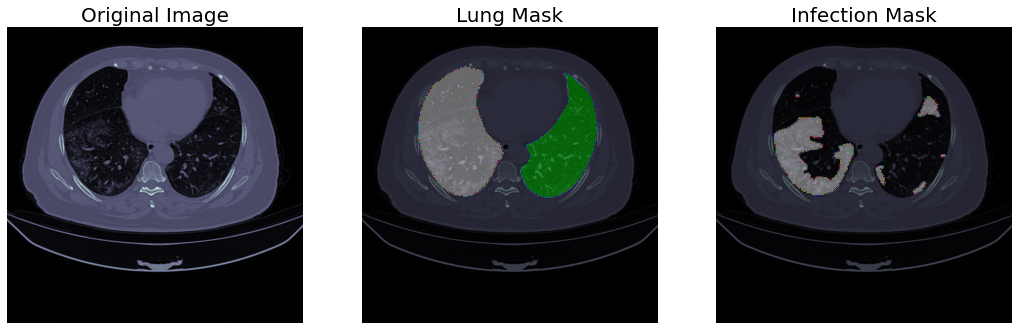}
    \caption{Example of CT-scan from the dataset $D_1$, with its corresponding lung and infection masks. }
    \label{fig:example_D1}
\end{figure}

%\subsubsection{Data Preparation}
%\subsubsubsection{Ct-scan Preprocessing}

Although the CT-scans in all three datasets are provided in NIfTI (Neuroimaging Informatics Technology Initiative) format, two datasets provide volumetric scans. In contrast, the last one provides axial CT slices (individual slices belonging to different patients). Therefore, single slices were extracted from volumetric CT-scans. For each volumetric CT-scan with COVID-19, the 20 middle slices were considered. The ground-truth infection masks were used to keep only infected slices.  For normal CT-scans, the 5 middle slices were extracted to keep a balanced dataset as the number of normal CTs is higher than that of COVID-19 cases.  Finally, all slices were converted to PNG (Portable Network Graphics) images, resized to 256*256 and their values normalized between 0 and 1. 

\begin{table}[h]
%\scriptsize
\small
\renewcommand{\arraystretch}{1.3}
\caption{Summary of the datasets used.}
    \centering
  %  \resizebox{\textwidth}{!}{
    \begin{tabular}{|c|c|c|c|c|} 

  \hline
  \multicolumn{1}{|c|}{\textbf{Dataset}} & \textbf{\# of Patients Used} & \textbf{\# of Images Used} & \textbf{Lung Mask} & \textbf{Lesion Mask}\\ 
  \hline
 %COVID-19 CT Lung and Infection Segmentation Dataset 
 $D_1$ & 20 Positive & 400 & \checkmark & \checkmark\\
  \hline
   %COVID-19 CT segmentation dataset  
   $D_2$& >40 Positive & 100 &\checkmark & \checkmark\\
  \hline
   %MosMedData 
   $D_3$ & 50 positive, 254 negative & 1770 & & \checkmark\\
  \hline
  \end{tabular}%}
    \label{tab:datasets}
\end{table}

Table \ref{tab:datasets} shows information about the used datasets.
%\subsubsection{Data augmentation}
To increase the data size and avoid the overfitting problem, helping thus in improving the classification performance, data augmentation was performed. The dataset was augmented by applying different transformation techniques such as horizontal flipping and rotation. %these images are considered separate images. %Examples of CT images after rotation and horizontal flipping are shown in Figures \ref{fig:ct_rotate} and \ref{fig:ct_flip}. After transformation, 

\subsection{Lung Segmentation}

To focus on the lung area, the lung segmentation step is considered. At this stage, U-Net architecture \cite{ronneberger2015u}, a well-known CNN designed for biomedical applications, is trained to generate lung masks for inputs (CT-scan slices). U-Net has achieved superior performance in many medical image segmentation applications, even with limited datasets. The primary aim of this segmentation step is to extract the lung regions (ROI) for the CTs in the $D_3$ dataset as it only provides infection masks. For this end,  $D_1$ and $D_2$ datasets are used to train the segmentation model. This latter takes individual CT slices as inputs and outputs the predicted lung masks. The segmented lung is later fed into the classification model for COVID-19 detection.

%Lung segmentation was performed using the well-known U-Net\cite{huang2017densely}, a CNN designed for biomedical applications. U-Net architecture is among the most popular CNN architectures for biomedical image segmentation. It has achieved state-of-the-art performance in many different medical image segmentation applications, even with limited datasets. %Example of a CT-scan after the lung segmentation is shown in Figure \ref{fig:seg_exp}. %before and after
%\begin{figure}[h]
 %   \centering
 %   %[width=0.8\textwidth]
 %   \includegraphics{images/20.JPG}
 %   \caption{Example of a CT-scan after the lung segmentation %
% step}
 %   \label{fig:seg_exp}
%\end{figure}

\subsection{COVID-19 Detection}

Convolutional Neuron Networks (CNNs) have been commonly used in many aspects of medical image analysis, and have contributed to the great progress in the diagnosis of rapidly evolving diseases. In this prediction component, we compared three different CNN architectures, namely standard CNN (custom) models, DenseNet, and ResNet, for COVID-19 detection. These architectures are briefly presented below.
\begin{itemize}
    \item ResNet architecture \cite{he2016deep} has contributed enormously to the use of very deep neural networks. It limits the gradient loss in the deepest layers by adding a residual bond between each convolution layer. Thus, their main advantages are their ability to reduce gradient leakage in the lower layers of the network, and the possibility of their deep scaling. The latter allows the adjustment of the architecture to the classification task. Although expensive in computational resources, they obtain very good performance in classification on all levels of image detail.

    \item DenseNet architecture \cite{huang2017densely} are in line with deep CNNs. Each of the network layers has as input the outputs of the previous layers. This feature increases the number of network connections from L, for a traditional L-layer CNN with one connection between each layer, to $[L(L-1)]/2$.  Although the DenseNet architecture adopts a similar approach to the ResNet architecture by adding short connections between the layers, it differs significantly in how information is taken from the previous layers. The concatenation of the information made in the DenseNet architecture makes it possible to reduce the number of parameters by a factor of 3 compared to the ResNet architecture while maintaining the same level of performance in terms of \cite{huang2017densely} precision.
    
    \item Our introduced classification model is a standard CNN. It consists of 4 convolution layers, each layer is followed by a max-poling layer. The role of these layers is the extraction of image features. After that, two dense layers and a Softmax function are used for classification based on the feature extracted from previous layers. Figure \ref{fig:CNN_arch} shows the CNN architecture considered for classification.

\end{itemize}

%In addition, we also used DenseNet \cite{huang2017densely} and ResNet \cite{he2016deep} architectures implemented in keras \cite{keras}  to extract the image features. After that, we used two dense layers and a Softmax function for classification based on the feature extracted by each architecture .\\

\subsection{Explainability of the model predictions}

% To make our framework more transparent and to provide the end-user with a better understanding of COVID-19 predictions, visualization as well as textual explanations are provided. 

To make our framework more transparent and increase the understandability of its decisions, visualization as well as textual explanations are provided.
The visual explanations contribute to understanding the model predictions by highlighting the regions in the input CT-scan that were behind the model's decision, helping to increase confidence in the model. As part of CT-xCOV, we use three different visual explainability techniques, namely LIME \cite{ribeiro2016should}, Grad-Cam \cite{selvaraju2017grad} and Integrated Gradients \cite{sundararajan2017axiomatic}. The goal is to make available to users different types (gradient-based and perturbation-based) of pixel attribution methods. The three used techniques are described below.

\begin{itemize}
    \item \textit{Local Interpretable Model-Agnostic Explanations (LIME)} is an XAI technique introduced in \cite{ribeiro2016should} to explain how an ML model input features affect its predictions. In the context of image classification, LIME identifies the region of an image, typically a set of super-pixels, that is most strongly associated with a predicted label. To generate explanations, %LIME creates a new dataset of randomly perturbed instances and their corresponding predictions, and fits a weighted local surrogate model. This surrogate model is usually a simpler, inherently interpretable model such as a linear regression model.
    the different steps of LIME are summarized in five steps; (1): Generate random perturbations for the input image. (2): Predict the class for each perturbation. (3): Compute importance weights for each perturbation. (4): Fit an explainable linear model using perturbations, predictions, and weights. (5): Explain the prediction by interpreting the interpretable model.
    \item \textit{Gradient weighted Class Activation Map (Grad-CAM)} \cite{selvaraju2017grad} is a visualization technique that is useful for understanding how a CNN was driven to make a classification decision. Grad-CAM uses target gradients in the last convolutional layer to provide explanations by highlighting important regions of an image that impact prediction. 
    %Grad-CAM is a variant of the CAM method \cite{cam} that, unlike CAM, considers gradients flowing in the last convolution layer not just the weights. The advantage of using gradients is that Grad-CAM can be applied to any layer of the network. 
    %This method consists of producing 
    The output of Grad-CAM is heatmaps representing the activation classes on the images received as input. An activation map is associated with a specific output class. The class activation map (CAM) is generated by calculating a linear combination of activations, weighted by the weights of the corresponding output for the observed class, This assigns importance to each position (x, y) in the image based on their contribution to the final prediction.
    \item \textit{Integrated gradients (IG)} is an explainable AI technique first introduced in \cite{sundararajan2017axiomatic}. The IG technique aims to provide explanations of how a model's predictions are influenced by its input characteristics. Its use cases include understanding feature importance, identifying data asymmetry, and debugging model performance. IG has become a popular interpretability technique due to its wide applicability to any differentiable model (text, images, structured data), and its ease of implementation. %In the case of images, calculating IG starts from the baseline, which can be a black, an all-white, or a random image, and generates a linear interpolation between this baseline and the original image. The interpolated images can be considered as small steps ($\alpha$) in feature space between the baseline and the input image, which constantly increases with the intensity of each interpolated image. Gradients are then calculated to measure the relationship between changes in a feature and changes in the model predictions. These gradients indicate which pixel has the strongest effect on the class probabilities predicted by the models. Finally, IG is scaled to the input image to ensure that attribution values are accumulated over multiple interpolated images in the same units. The IG is represented on the input image with the pixel importance. The formula for integrated gradients is: $$IntegratedGradients_{i}(x)::= (x_{i}-x'_{i})  \int_{\alpha = 0}^{1} \dfrac{\partial F(x'+\alpha(x-x'))}{\partial x_{i}} \, \mathrm{d}\alpha $$ \\Where: \\ i: feature (individual pixel).\\ x: input (image).\\ x' : baseline (image).\\ k: scaled characteristic disturbance constant.\\ m: number of steps in the Riemann sum approximation of the integral.\\ $(x_{i}-x'_{i})$: a term for the difference from baseline.\\
\end{itemize}

In addition to visualization, CT-xCOV also provides textual explanations to report information about the degree of lung infection. It is computed based on the explanation provided by the visualization methods. The degree of infection is calculated as the size of the infected region, highlighted by the visualization, divided by the size of the lungs.

%\subsubsection{\textcolor{purple}{Explanations assessment}}

\section{Experiments}
\label{exp}

To assess the effectiveness of our CT-xCOV framework, we performed several experiments with different settings. This section describes the experimental setup and defines the metrics used for evaluation.

%In this section, we introduce our experimental settings as well the considered evaluation metrics used in the CT-xCOV framework
%We start with the results obtained for the segmentation step followed
%by the classification results. After that, we discuss the explanation of each method
%separately. Finally, the results of the comparison between these explanations are discussed.

\subsection{Experimental setup}

%\textcolor{red}{FIXME:il faut compléter had la partie avec les détails sur l'entrainement u-net stp: For the segmentation task in the proposed framework, we trained the two dataset, namely   using the U-Net \cite{} ....}

\paragraph{Segmentation task.}

For the lung segmentation component in the proposed framework,  we adopted the basic architecture of U-Net \cite{ronneberger2015u}, with the difference in the input shape becoming (256*256) and the output shape becoming (256*256). U-NET was trained using 554 CT-scan images of 60 patients, and 200 images of 10 patients were kept for testing.  Since no lung masks are provided in $D_3$, the images used for the segmentation task were mainly taken from  $D_1$ and $D_2$ datasets. To increase the number of training images, some slices were segmented using the pre-trained model provided by the MedSeg \cite{MedSeg} tool. To ensure that the test images are not seen during the training process of U-Net,  we performed a train/test split at the patient level. Thus, images from 10 patients were kept for testing. Finally, the training hyper-parameters are summarized in Table \ref{tab:segmentation_param}.

\paragraph{Classification task.}

\begin{figure}[h]
    \centering
        \includegraphics[width=0.9\textwidth]{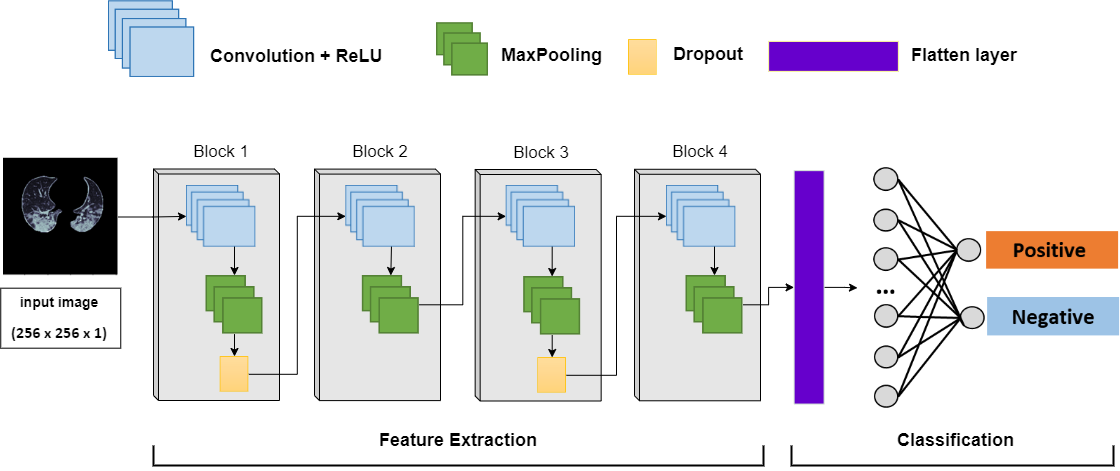}
    \caption{The architecture of our classification model}
    \label{fig:CNN_arch}
\end{figure}

We evaluated this task using three models. On the one hand, we used DenseNet121 \cite{DBLP:journals/corr/HuangLW16a} and ResNet50 \cite{DBLP:journals/corr/HeZRS15,} architectures to extract the image features. Two dense layers and a Softmax function are then employed for classification based on the feature extracted by each architecture. On the other hand, we designed the simple CNN architecture of Figure \ref{fig:CNN_arch}, constituted of 4 layer blocks for feature extraction and two dense layers for classification. Each layer block is composed of  1) a convolution layer with 32 filters of $3\times3$ and ReLu activation function followed by 2) a max pooling layer. For model regularization, a dropout layer is added to the first and last blocks. To ensure that the test images were not seen during the training process, we performed a train/test split at the patient level, where 15\% of the images belonging to 50 patients were kept for the test. The training hyper-parameters are summarized in Table \ref{tab:segmentation_param}.

Finally, classification and segmentation models were implemented using Keras \cite{keras}  and Tensorflow libraries with Python 3.7 on  Colab (NVIDIA Tesla K80 GPU, 12 GB RAM) and Kaggle (NVIDIA Tesla P100 GPU, 16 GB RAM).   

\paragraph{Explanation assessment.}
%\subsubsection{\textcolor{purple}{Infection segmentation based on visual explanation }}
The goal of this experiment is to compare and assess the explanations provided by the visualization methods. To this end, we adopt a ground-truth-based evaluation, where the visualization output is compared to the ground-truth infections. The comparison is performed by measuring the similarity between the two images. For each CT-scan, %visual explanations are generated using the LIME, Grad-Cam, and Integrated gradients, and compared with the ground-truth infections to measure the similarity coefficient. %The performance of the three visualization methods is assessed using the average of the computed similarity coefficient.
we measured the similarity between the output explanations and the ground-truth infection masks. In fact, for the LIME method, we consider the output part of the explanation as the infected region while for heatmap methods (Grad-cam and Integrated Gradient), we applied threshold-based segmentation to extract the highlighted infections.  Image thresholding is a well-known image binarization technique, which consists of Converting a grayscale image into a binary image, where the pixel values are restricted to either 1 or 0. Since this thresholding technique depends mainly on the threshold value, different threshold values were tested using the train set to find the value that gives the best segmentation results compared to the infection masks.

\begin{table}[h]
\small
\renewcommand{\arraystretch}{1.1}
    \caption{Hyper-parameters values used in segmentation and classification model}
    \centering
    \begin{tabular}{ | l | l | l |} 
  \hline %\rowcolor{lightgray}
  \textbf{Hyper-parameters} & \textbf{Segmentation Model} & \textbf{Classification Model} \\ 
  \hline
  Batch Size & 32 & 32\\ 
  \hline
   Epoch & 100 & 20\\ 
  \hline
  Learning rate & 0.0005 & 0.001\\ 
  \hline
   Optimizer & Adam & Adam\\ 
  \hline
   Loss & Binary Cross Entropy and Dice Loss & Categorical Cross Entropy\\ 
  \hline
  \end{tabular}
    \label{tab:segmentation_param}
\end{table}

\subsection{Evaluation metrics}

The evaluation of CT-xCOV is performed according to the three considered tasks: lung segmentation, classification, and visual explanations.

\paragraph{Classification metrics.} 
We considered different evaluation metrics to report the performance of the evaluated models, namely precision, recall, accuracy, false discovery rate, and F1-score. 
For classification, prediction outcomes belong to one of the four cases: (*) True Positive (TP) corresponds to positive samples that are correctly classified as positive, (*) True Negative (TN) indicates samples that are actually negative and are correctly classified as negative, (*) False Positive (FP) refers to negative samples incorrectly classified, and (*) False Negative (FN) refers to positive samples incorrectly classified. Based on these outcomes, evaluation metrics are defined as given below:

\begin{itemize}%[noitemsep,nolistsep]
    \item \textbf{Accuracy}: is a measure of correct predictions over total predictions:
    $$Accuracy=\frac{TP+TN}{N}$$

    \item \textbf{Precision}: measures the proportion of TP among the samples classified as positive:
    $$Precision=\frac{TP}{TP+FP}$$
    \item \textbf{Recall}: measures the proportion of actual positives that are correctly identified:
    $$Recall=\frac{TP}{TP+FN}$$
    \item \textbf{$F_{1 }$ score}: is a weighted average of Precision and Recall: 
    %It takes both false positives and false negatives into account. 
    $$F_{1 }score =2 \times \frac{Recall \times Precision}{Recall+Precision}$$
    \item \textbf{False discovery rate }: is the portion of FP among all positive predictions:  $$FDR=\frac{FP}{FP+TP}$$
\end{itemize}

\paragraph{Segmentation and explanations assessment metrics.}
Dice coefficient \cite{haque2020deep} was used for these tasks. It is a widely adopted statistical indicator used to quantify the similarity between two sample images. Dice measures the spatial overlap between two images A and B as given by the formula below, where the cardinalities refer to the number of pixels. Alternatively, Dice can be expressed in terms of TP, FP, and FN, which refers to the number of pixels in A that are classified correctly or incorrectly concerning the B.
$$Dice(A, B) =\frac{2 |A \cap B|}{|A|+|B|} =\frac{2 \times TP}{2 \times TP+FP+FN} $$ 

%\textcolor{red}{FIXME: a vérifier avec Pr Afaf ref\cite{haque2020deep}}

\section{Results}
\label{res}
This section discusses the results obtained at each step
of our CT-xCOV framework, namely segmentation, classification, and explainability.

\subsection{Lung segmentation}
\begin{figure}[h]
    \centering
    \includegraphics[width=0.5\textwidth]{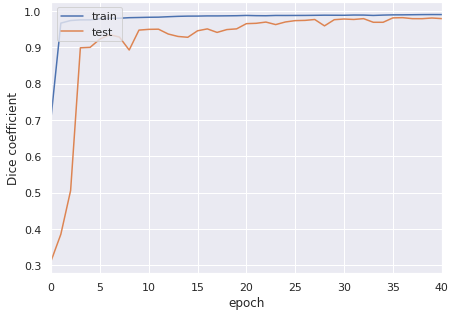}
    \caption{Dice coefficient curves for the U-Net training and test sets.}
    \label{fig:unet_acc}
\end{figure}

The Dice coefficient curves for the training and validation of U-Net during the segmentation step are illustrated in Figure \ref{fig:unet_acc}.  Comparing the two curves, it can be noticed that, as the training progresses, the test Dice coefficient follows the train Dice coefficient while remaining slightly inferior. This means that the model's performance increases over time, so the model is learning. The model achieves a value of 98.33\% on the test set.

\begin{figure}[h]
    \centering
    \includegraphics[width=0.8\textwidth]{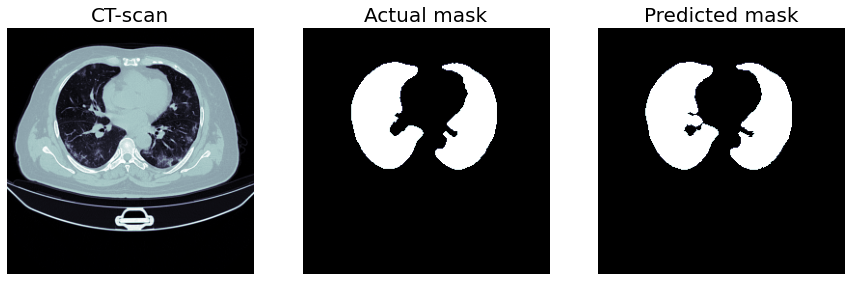}
    \caption{Example of current mask and mask predicted by U-Net}
    \label{fig:my_unet_pred}
\end{figure}

A visual example of U-Net predictions is shown in Figure \ref{fig:my_unet_pred}, which compares an output mask of a random image with the ground-truth mask. We can see that the predicted mask is highly consistent with the ground-truth.
%\vspace*{-0.5cm}

\subsection{Classification for COVID-19 Detection}

The comparison between our proposed CNN model and the other architectures (Resnet and DenseNet) was performed using the 5-fold cross-validation strategy on the train set.
The cross-validation results are summarized in Table \ref{tab:comp_arch1}. It can be observed that the best results are achieved with our proposed CNN. An other difference can be seen in the stability of the models. The performance scores of ResNet50 and DenseNet121 have higher variance (between 3.7\% to 12.7\%), which means that the predictions of the trained models are unstable. On the other hand, our model achieved higher performance with low variance ($\leq$ 1\%).

\begin{table}[h!]
\small
\renewcommand{\arraystretch}{1.1}
    \caption{Comparison between our model and the other architectures.}
    \centering
    \begin{tabular}{ | l| l | l| l | l |} 
  \hline %\rowcolor{lightgray}
   \textbf{Model}  & \textbf{Precision} & \textbf{Recall} & \textbf{F1-score} & \textbf{Accuracy} \\ 
  \hline
   Our Model &\textbf{ 98.93$\pm$0.37\%}   & \textbf{97.54$\pm$0.99\%}  & \textbf{98.23$\pm$0.46\%}  & \textbf{98.40$\pm$0.41\%} \\ 
  \hline
  ResNet50 & 82.31$\pm$14.08\%   & 92.32$\pm$2.63\% & 86.34$\pm$ 9.22\% & 85.30$\pm$12.60\%\\ 
  \hline
  DenseNet121 & 98.42$\pm$ 1.92\% & 69.34$\pm$ 17.69\% & 80.03$\pm$11.34\% & 85.44$\pm$7.35\%\\ 
  \hline
  \end{tabular}
    \label{tab:comp_arch1}
\end{table}
The model performance on the test dataset is summarized in Table \ref{tab:comp_arch2}. %Although we can see that DenseNet121 achieved higher performance (84\%), these values are unreliable since the model has been proven to be unstable. Note that, a lower performance has been achieved by ResNet50.
%
%As the values in Table \ref{tab:comp_arch2} show, 
Our model achieved good performance on the test set, with higher scores of accuracy (93\%), precision (93\%), recall (94\%), F1-score (93\%), and a lower score of False discovery rate (11.5\%). 
\begin{table}[h!]
    \centering
    \small
    \renewcommand{\arraystretch}{1.1}
    \caption{Comparison between our model and other CNN architecture on the test dataset }
    \begin{tabular}{ | l | l | l | l | l | l |} 
  \hline %\rowcolor{lightgray}
   \textbf{Score}  & \textbf{Precision} & \textbf{Recall} & \textbf{F1-score} & \textbf{Accuracy} & \textbf{FDR}\\ 
  \hline%\rowcolor{yellow!50}
   Our Model &\textbf{ 93\% }& \textbf{94\%}& \textbf{93\%} & \textbf{93\%} & \textbf{11.5\%} \\ 
  \hline
  ResNet50 & 77\%   & 77\% & 76\% & 76\% & 29.5\%\\ 
  \hline
  DenseNet121 & 84\%   & 83\% & 82\% & 82\% & 32.5\%\\ 
  \hline
  \end{tabular}
        \label{tab:comp_arch2}
\end{table}

\subsection{Explainability of the model predictions}
%\subsubsection{Visual explanations}
To gain insights into the outcomes of our models, we aim to create visual representations of the results. we generate visual explanations using different explainability techniques, namely Grad-CAM, Integrated Gradient (IG), and LIME.

\begin{figure}[h]
\centering
\begin{subfigure}{0.3\textwidth}  
 \centering
    \includegraphics[width=0.7\textwidth]{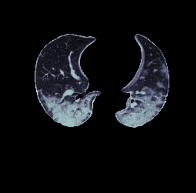}
    \caption{CT-scan}
    \label{fig:ct}
\end{subfigure}\hspace{-10mm}
\begin{subfigure}{0.3\textwidth}  
 \centering
    \includegraphics[width=0.7\textwidth]{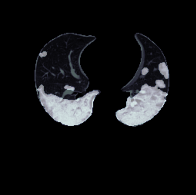}
    \caption{Ground truth for infection}
    \label{fig:infec}
\end{subfigure}
\\
\centering
\begin{subfigure}{0.3\textwidth}  
 \centering
    \includegraphics[width=0.7\textwidth]{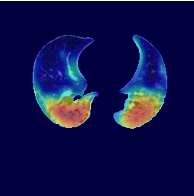}
    \caption{Grad-CAM }
    \label{fig:my_gradcam1}
\end{subfigure}\hspace{-10mm}
\begin{subfigure}{0.3\textwidth}  
 \centering
    \includegraphics[width=0.7\textwidth]{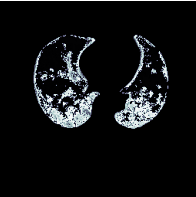}
    \caption{Integrated Gradient  }
    \label{fig:IG1}
\end{subfigure}\hspace{-10mm}
\begin{subfigure}{0.3\textwidth}
    \centering
    \includegraphics[width=0.7\textwidth]{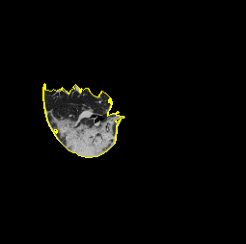}
    \caption{LIME}
    \label{fig:lime1}
    \end{subfigure}
    \caption{Visualizing positive COVID-19 CT-scan}
    \label{fig:XIAcov}
\end{figure}

Figure \ref{fig:XIAcov} illustrates an example of explanations produced by the three techniques for COVID-19-positive cases, together with the ground-truth infection mask. Comparing the output of the methods, we notice that: 
\begin{itemize}
    \item Grad-CAM (Figure \ref{fig:my_gradcam1}) correctly highlighted the infected parts of the lungs in the CT-scan, which provides an understanding of the model predictions.
  \item IG (Figure \ref{fig:IG1}) correctly localized a wide part of COVID-19 infection in the image.
    \item LIME (Figure \ref{fig:lime1}) with the top two features obtained using 1000 perturbations,  highlighted the regions corresponding to the infected lung area. However, we can see that LIME localized only a part of the infection.

\end{itemize}

\begin{figure}[h!]
\centering
\begin{subfigure}{0.3\textwidth}  
 \centering
    \includegraphics[width=0.7\textwidth]{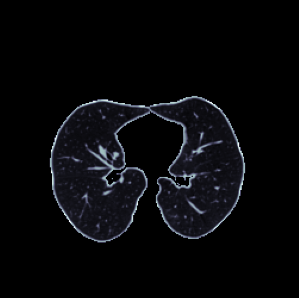}
    \caption{CT-scan}
    \label{fig:ct2}
\end{subfigure}\hspace{-10mm}
\begin{subfigure}{0.3\textwidth}  
 \centering
    \includegraphics[width=0.7\textwidth]{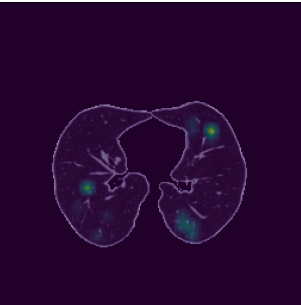}
    \caption{Grad-CAM visualization}
    \label{fig:gradcam2}
\end{subfigure}

\centering
\begin{subfigure}{0.3\textwidth}  
 \centering
    \includegraphics[width=0.7\textwidth]{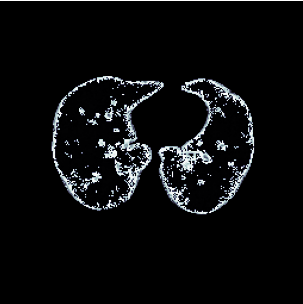}
    \caption{Integrated Gradient  }
    \label{fig:IG2}
\end{subfigure}\hspace{-10mm}
\begin{subfigure}{0.3\textwidth}
    \centering
    \includegraphics[width=0.7\textwidth]{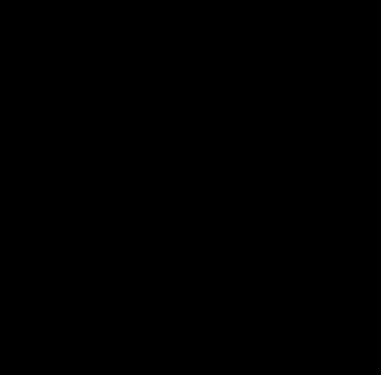}
    \caption{LIME  }
    \label{fig:lime2}
    \end{subfigure}
    \caption{Visualizing negative  COVID-19 CT-scan}
    \label{fig:XIAnocov}
\end{figure}

Figure \ref{fig:XIAnocov} illustrates an example of explanations produced by the three techniques for COVID-19 negative cases, together with the ground-truth infection mask.  Comparing the output of the methods, we notice that:

\begin{itemize}
    \item Grad-CAM (Figure \ref{fig:gradcam2}) shows no sign of infection which explains why the model predicted the CT-scan as negative.
    \item IG (Figure \ref{fig:IG2} highlights the contour of the lungs, which may not be as good as other methods' visualizations.
    \item LIME (Figure \ref{fig:lime2}) outputs a black image, which indicates that no infection has been detected.
    
 %% negative case
\end{itemize}

%The examples of visual explanations for the negative COVID-19 CT-scan are illustrated in Figure \ref{fig:XIAnocov}. 

%%%%%%%%%%%%%%%%%%%%%%%%%%%%%%%%%%

\begin{figure}[h]
    \centering
    \includegraphics[width=0.99\textwidth]{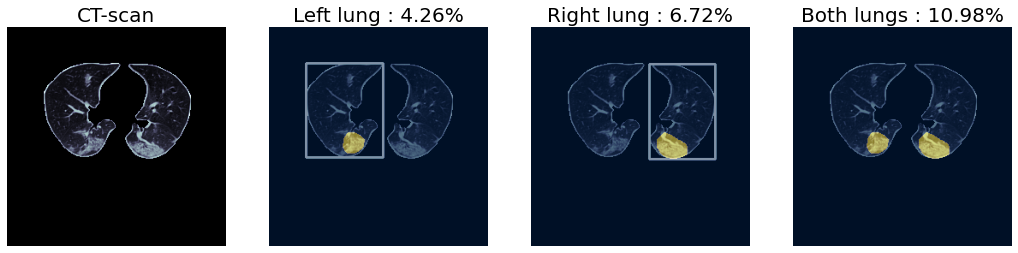}
    \caption{Percentage of infection in the lungs based on the explanation provided by Grad CAM; "Total degree of infection in both lungs:  10.98 \%, 4.26 \% in the left lung and 6.72 \% in the right lung".}
    \label{fig:text_exp}
\end{figure}

As mentioned before, CT-xCOV enriches the visualization's explanations with the textual ones. We provide users with information on the degree of infection, which is calculated as the size of the infected region shown by a visualization method divided by the size of the lungs. Additionally, we provide the degree of infection per lung (i.e. right/left) to give more information about the most infected lung.  Figure \ref{fig:text_exp} shows an example of visual and textual explanations using the Grad-Cam method:  the infected region is highlighted and the percentage of infection is computed.

\subsection{Explanations assessment}
\begin{figure}[h]
    \centering
    \includegraphics[width=\textwidth]{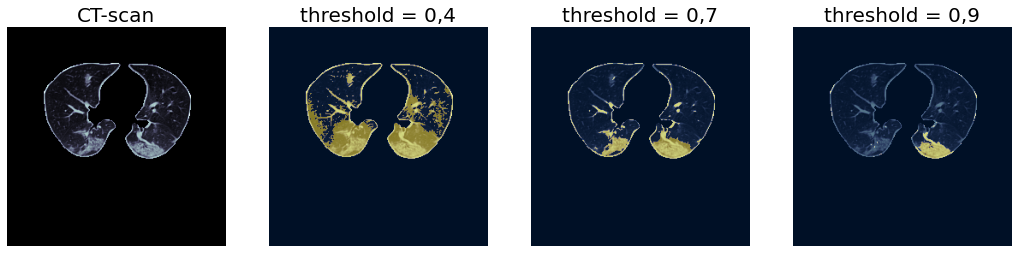}
    \caption{The impact of threshold choice on grad-Cam's explanation}
    \label{fig:treshold_com}
\end{figure}

As we mentioned before, threshold-based segmentation has been used for segmenting the output visualization. This segmentation depends on the threshold value as illustrated in Figure \ref{fig:treshold_com}, which shows the infection segmentation obtained by Grad-CAM using different threshold values.

\begin{figure}[h!]
    \begin{subfigure} {0.49\textwidth}
    \centering
    \includegraphics[width=1\textwidth]{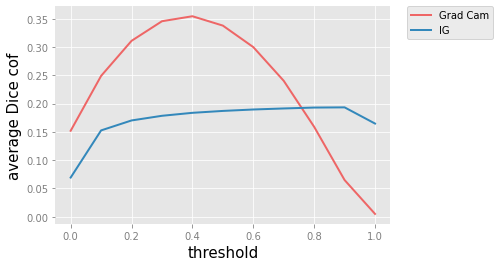}
    \caption{COVID-19 positive}
    \label{fig:tresh_d_cov}
\end{subfigure}
\begin{subfigure}{0.49\textwidth}
    \centering
    \includegraphics[width=1\textwidth]{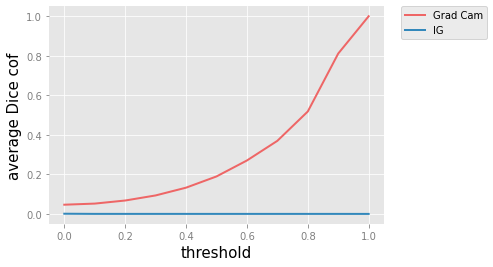}
    \caption{COVID-19 negative}
    \label{fig:tresh_d_nocov}
    \end{subfigure}
    \caption{Impact of threshold value on the Dice coefficient for COVID-19 positive and negative cases.}
    \label{fig:thresholding}
\end{figure}

Given the difference between the visualizations provided by the methods in the case of NO-COVID images (IG highlights also the lung contour as a part of the visualization (see Figure \ref{fig:XIAnocov})), we decided to evaluate the methods for the COVID-19 positive and negative cases separately. Figure \ref{fig:thresholding} shows the dice coefficient plots for the tested threshold values.
The Dice coefficients for different threshold values, on COVID-19 positive cases are plotted in Figure \ref{fig:tresh_d_cov}. Comparing the plots of the different methods, we can see that the Grad-cam achieved the best result of 0.35 with a threshold of 0.4, followed by the Integrated gradient which achieved 0.19 for a threshold of 0.9.
The plots for COVID-19 negative cases are shown in Figure \ref{fig:tresh_d_nocov}. Comparing the plots, it can be observed that Grad-Cam achieved an average Dice coefficient of 0.81 with a threshold of 0.9, followed by Integrated Gradient for which the average Dice coefficient did not exceed 0.00024 with the threshold of 0.1.  
Table \ref{tab:exp_comp1} summarizes the comparison findings for the threshold choice. % that give the better infection segmentation.

\begin{table}[h]
\small
\renewcommand{\arraystretch}{1.1}
    \centering
    \begin{tabular}{ | l | l | l| l| l| l|} 
  \hline 
   \multirow{2}{*}{\textbf{XAI Method} } & \multicolumn{2}{c|}{\textbf{COVID-19 positive cases}} & \multicolumn{2}{c|}{\textbf{ COVID-19 negative cases}}\\
   \cline{2-5}
   & \textbf{Threshold} & \textbf{Average Dice coeff} & \textbf{ Threshold} & \textbf{Average Dice coeff} \\ 
  \hline
  Grad-cam & 0.4 & 0.35 & 0.9 & 0.81  \\ 
  \hline
  Integrated Gradient & 0.9 & 0.19  & 0.1 & 0.00024  \\
  \hline
  %LIME & - & 0.24 & - & 0.77  \\ 
  %\hline

  \end{tabular}
    \caption{Best threshold values found on the train set for segmenting Grad-CAM and IG visualizations}
    \label{tab:exp_comp1}
\end{table}

After defining the infection segmentation thresholds in the visualization images, the comparison between the XAI methods is performed on the test dataset, using the average of the Dice coefficient as the metric. The results
are summarized in the table \ref{tab:assessemnt}.

\begin{table}[h!]
    \centering
    \small
    \renewcommand{\arraystretch}{1.1}
    \caption{Summary of explanations assessment based on the test dataset}
    \begin{tabular}{ | l | l | l | l | } 
  \hline %\rowcolor{lightgray}
   \textbf{XAI Method}  & \textbf{Average Dice coefficient (Covid case )} & \textbf{Average Dice coefficient
(Normal case )} & \textbf{Explanation Time} \\ 
  \hline
   Grad-Cam & \textbf{55\%} & \textbf{65\%} & \textbf{0.33s}\\ 
  \hline
  Integrated Gradient & 29 \%   & 0.016 \%  & 0.41s\\ 
  \hline
  LIME & 24 \%   & 77  \% & 18s\\ 
  \hline
  \end{tabular}
        \label{tab:assessemnt}
\end{table}

The summarized Dice coefficient scores show that:

\begin{itemize}
    
    \item For COVID-19 positive scans, Grad-CAM visualizations significantly outperform IG and LIME visualizations by achieving a 55\% overlapping with the ground-truth masks versus 29\% and 24\% achieved by IG and LIME respectively.
    
    \item  For COVID-19 negative scans, except LIME, higher Dice coefficient values were obtained since we compared visualizations with black images (meaning no lung infection). The low value obtained for IG can be explained by the fact that the IG method produces lung contours leading to high dissimilarity when compared to the black masks.
    
\end{itemize}

An additional comparison between the visualization methods was also made by considering the average explanation time (i.e. the time taken to produce the visual explanations). The results obtained are also presented in Table \ref{tab:assessemnt}. These results show that LIME  is time-consuming (an average time of 18s), compared to Grad-CAM and IG which give explanations in less than 1s.

%%
%%\begin{table}[h]
%%\renewcommand{\arraystretch}{1.1}
%%\small
%%    \centering
%%    \begin{tabular}{ | l | l | l | l |} 
%%  \hline %\rowcolor{lightgray}
%%   \textbf{XAI Method}  & Grad-Cam & Integrated Gradient & LIME  \\
%%  \hline
%%  \textbf{Average Explanation Time} & 0.33s & 0.41s & 18s  \\ 
%%  \hline
%%  \end{tabular}
%%    \caption{ average explanation calculation time for each method}
%%    \label{tab:exp_comp3}
%%\end{table}

\section{Related works}\label{relWork}
This section provides an overview of previous research related to COVID-19 diagnosis. It is divided into two main subsections. In the first subsection, existing approaches for COVID-19 diagnosis based on medical imaging are presented. The second subsection investigates existing works on explainable artificial intelligence methods for COVID-19 diagnosis. %The main findings of these works are summarized in Table\ref{tab:summary}.

\subsection{COVID-19 detection based on medical imaging}

Several papers propose approaches for detecting COVID-19 infection based on medical imaging. These papers use different techniques, some use DL models  (\cite{anwar2020deep,ahsan2020study,he2020sample}) and others combine image processing techniques with machine learning models (\cite{khan2020covid,fauvel2006combined}).  

Researchers, recently, are most interested in applying DL techniques for COVID-19 diagnosis. The authors in \cite{anwar2020deep} used a CNN architecture called EfficientNet-b4 to distinguish between infected and normal CT-scan images. The dataset used in this paper consists of 349 CT-scan images of 216 COVID-19 patients and 396 images of non-COVID patients. The dataset was augmented using different transformation techniques such as horizontal/ vertical flipping, and rotation; images were also shifted left, right, upward, and downward. The model performance was evaluated using 5-fold cross-validation. The trained model achieved an Accuracy of 90\% and an F1-score of 90\%.
In another paper \cite{ahsan2020study}, the authors aim to accurately detect COVID-19 infection on two types of imaging datasets (CT-scan and chest X-ray). To this end, different DL models such as InceptionResNetV2, VGG16, ResNet50, MobilenetV2, NasNetMobile, DenseNet201, VGG19, and ResNet15V2 have been evaluated. %Each dataset contains 200 COVID cases and 200 normal cases. 
The results demonstrate that NasNetMobile exhibited superior performance in terms of accuracy compared to all other models on both datasets (with scores (81.5\% - 95.2\%) for CT-scan and (95.4\% - 100\%) for chest X-ray).
In \cite{alshazly2021explainable}, the authors evaluated several DL architectures for COVID-19 diagnosis. Their proposed approach involved utilizing transfer learning with custom-sized inputs optimized for each specific architecture to attain optimal performance. There are several models included in this paper, such as SqueezeNet, Inception, ResNet, Xception, and DenseNet. They performed a series of experiments on two datasets of CT-scan images, namely, the  SARS-CoV-2 CT-scan and COVID-19. The study reported that the evaluated models outperformed previous studies.
On the SARS-cov-2 CT dataset, the models achieved a score of 99.4\% and 99.8\% for accuracy and sensitivity respectively. The scores achieved on the COVID-19 datasets are 92.9\% for accuracy and 93.7\% for sensitivity.

In \cite{el2021automated} the authors involved creating optimized versions of various CNN architectures, including VGG16, VGG19, DenseNet201, Inception\_ResNet\_V2, Inception\_V3, Xception, Resnet50, and MobileNet\_V2, through fine-tuning. To prevent overfitting within their models, they used weight decay and regularizers, known as L2.  The performance of these models was evaluated using a dataset comprised of chest X-ray and CT scans, for binary classification. According to the findings, Resnet50, MobileNet\_V2, and Inception\_Resnet\_V2 which were fine-tuned demonstrated exceptional performance with a considerable increase in both training and validation accuracy, achieving more than 96\% accuracy. However, CNN, Xception, VGG16, VGG19, Inception\_V3, and DenseNet201 had relatively low performance, with an accuracy of no more than 84\%.

Other works combine image processing techniques with machine learning models for COVID-19 detection.
Authors of \cite{khan2020covid} use Histogram of Oriented Gradients (HOG) to extract features in chest X-ray images of COVID-19. They compared the accuracy of several machine learning classification models, namely SVM, KNN, NB algorithm, and DT. The results demonstrated that the SVM is more accurate than other machine learning classification models. In \cite{fauvel2006combined}, the authors apply two DL models (DenseNet and GoogleNet) for feature extraction and stacking ensemble learning technique to  COVID-19 diagnosis, using an X-Ray dataset. The result indicated that the final accuracy of the proposed model was good compared to previous work.

\subsection{Explainable AI for COVID-19 Diagnosis}
Lately, much research on automatic COVID-19 diagnosis has been oriented toward creating human-understandable explanations of how DL models make specific individual predictions.
In \cite{ahsan2020study}, authors use LIME to explain their classification model.
A weighted linear model was employed, which determined a coefficient for each super-pixel in the image. These coefficients indicate the level of impact that each super-pixel has on the prediction of COVID-19 infection. After this, the top features were ranked to identify the most significant super-pixels.
In \cite{alshazly2021explainable}, The authors utilized two visualization techniques to offer visual explanations of the predictions made by their models, namely t-distributed stochastic neighboring embedding (t-SNE) and Grad-CAM.
The visual representations suggest that the models accurately pinpoint the areas linked to COVID-19.
In \cite{ye2021explainable}, the authors propose an explainable COVID-19 classifier for CT-scan dataset. The proposed approach includes an XAI module that can offer supplementary diagnostic information to radiologists. To perform their approach, CT-scan data was collected from hospitals in China. The collected data consists of 380 CT-scan with COVID-19 infection and 424 negative CT-scan. To ensure a just evaluation, models were trained on a private dataset and then tested on the open access publicly available CC-CCII dataset, which contains 2034 CT-scan. The proposed XAI module provides visualization based on CAM (Class Activation Map). Moreover, The adoption of the LIME and Shapley methods was aimed at interpreting the individual contributions of each super-pixel and examining the differences between local and global explainability. By utilizing these XAI modules, it was possible to identify the areas within the CT images that are more indicative of lesion presence, thereby enabling the model to arrive at a clear final decision.

%Contribution

The above discussion of related research demonstrates the crucial role of artificial intelligence in addressing the challenge of automated COVID-19 detection, as well as the increasing interest in interpretability/explainability to provide trustworthy detection systems.
Table \ref{tab:summary} presents a summary of the discussed works. For the sake of simplicity, for the works evaluating several DL models, we reported the results of the best model.
%the most relevant related works that adopted Artificial intelligence and explainable machine learning techniques to COVID-19 diagnosis .
Despite the success of ML and DL methods, which have proven to be very powerful for medical image classification, and the introduction of XAI methods, challenges in evaluating the effectiveness of XAI methods and in providing understandable explanations still remain. 

\begin{table}[h]

\caption{Comparative table of most relevant related works.}
    \centering

  \begin{tabular}{ccccc}

  \hline
  \textbf{Ref} & \textbf{Dataset} & \textbf{Classification approach} & \textbf{Performance} & \textbf{Interpretation} \\ 
  \hline
   \multirow{2}{0.7cm}{ \cite{anwar2020deep}}  & \multirow{2}{4.2cm}{COVID-CT-Dataset \cite{yang2020covid}} &  \multirow{2}{3.5cm}{EfficientNet b4 \cite{tan2019efficientnet}} %to discriminate between COVID CT-scan images and normal CT-scan images
   & %5 fold cross-validation:
   - Accuracy=90\%
   & \multirow{2}{2.5cm}{$-$} \\
   &&&- F1-score=90\% &\\
  \hline
   \multirow{3}{0.7cm}{ \cite{ahsan2020study}} & \multirow{3}{4.2cm}{CT and X-ray images collected from Kaggle}
 &  \multirow{3}{3.5cm}{NasNetMobile}
 & Accuracy on: 
  & \multirow{3}{2.5cm}{LIME}\\
 &&&  - CT (81.5\%–95.2\%)&\\
 &&&  - X-ray (95.4\%–100\%)&\\
   
   \hline
   \multirow{2}{0.7cm}{ \cite{khan2020covid}} & 
   \multirow{2}{4.2cm}{X-ray images  \cite{cohen2020covid}}  & \multirow{2}{3.5cm}{XGBoost}   & - Acc=97.87\% & \multirow{2}{2.5cm}{$-$} \\
   & &  & - F1-score=97.87\%&\\
   \hline
   \multirow{4}{0.7cm}{ \cite{alshazly2021explainable}} & 
   \multirow{2}{4.2cm}{SARS-CoV-2-CT-scan \cite{medicalsegmentation}} &\multirow{2}{3.5cm}{ResNet101} & - Acc=99.4\%&\multirow{4}{2.5cm}{Grad-CAM}\\
   & & &- F1-score=99.4\%&\\
   \cline{2-4}
   & \multirow{2}{4.2cm}{COVID19-CT  \cite{he2020sample}} & \multirow{2}{3.5cm}{DenseNet} & - Acc=92.9\%&\\
   &&&- F1-score=92.9\%&\\
   \hline
   \multirow{2}{0.7cm}{\cite{ye2021explainable}} & \multirow{2}{4.2cm}{CC-CCII dataset \cite{zhang2020clinically}} & \multirow{2}{3.5cm}{XAI based classifier} & Acc=89.23\%& \multirow{2}{2.5cm}{CAM, LIME and SHAP}\\
  &&&AUC=93.22\%& \\
  
   \hline
  \end{tabular}
    
    \label{tab:summary}
\end{table}

%\clearpage
\section{Conclusion}\label{conc}

In this paper, we propose CT-xCOV, an end-to-end approach for explainable COVID-19 diagnosis using CT-scans. The framework consists of four main components: (1) Data preparation, (2) lung segmentation, (3) COVID-19 detection, and (4) Explainability. 

The proposed CNN manages to achieve accuracy and an F1-score of ~98\% and is more stable compared to the two evaluated models. 
The visual explanations, obtained by the different XAI methods tested, show that the proposed CNN model has correctly detected COVID-19-related regions as specified in the infection masks. Such explanations are very useful to create a relationship of trust between the DL model and the experts in the field. 
Given the lack of standard metrics for evaluating XAI methods, we considered a ground-truth-based evaluation method using infection masks to compare the visualization performance of these methods. This evaluation consists of comparing and measuring the similarity between generated visualization and ground-truth infection masks using the well-known Dice coefficient metric. This comparison shows that Grad-CAM achieves good results in terms of the Dice coefficient (56\% on positive cases, and 65\% on negative cases) and explanation time (0.33s).
In future work, we intend to implement and investigate other metrics for evaluating interpretability, which will make a huge contribution to the growth of this line of research. We also intend to compare explainability achieved through DL-based approaches with
approaches using interpretable ML approaches combined with feature extraction. The aim of this is to create clear and defined features from CT-scan images using traditional image analysis techniques.

\printbibliography

@article{sun2021deep,
  title={Deep learning model improves radiologists’ performance in detection and classification of breast lesions},
  author={Sun, Yingshi and Qu, Yuhong and Wang, Dong and Li, Yi and Ye, Lin and Du, Jingbo and Xu, Bing and Li, Baoqing and Li, Xiaoting and Zhang, Kexin and others},
  journal={Chinese Journal of Cancer Research},
  volume={33},
  number={6},
  pages={682},
  year={2021},
  publisher={Beijing Institute for Cancer Research}
}

@article{kwee2020chest,
  title={Chest CT in COVID-19: what the radiologist needs to know},
  author={Kwee, Thomas C and Kwee, Robert M},
  journal={Radiographics},
  volume={40},
  number={7},
  pages={1848},
  year={2020},
  publisher={Radiological Society of North America}
}

@article{DBLP:journals/corr/HuangLW16a,
  author    = {Gao Huang and
               Zhuang Liu and
               Kilian Q. Weinberger},
  title     = {Densely Connected Convolutional Networks},
  journal   = {CoRR},
  volume    = {abs/1608.06993},
  year      = {2016},
  url       = {http://arxiv.org/abs/1608.06993},
  eprinttype = {arXiv},
  eprint    = {1608.06993},
  bibsource = {dblp computer science bibliography, https://dblp.org}
}

@article{DBLP:journals/corr/HeZRS15,
  author    = {Kaiming He and
               Xiangyu Zhang and
               Shaoqing Ren and
               Jian Sun},
  title     = {Deep Residual Learning for Image Recognition},
  journal   = {CoRR},
  volume    = {abs/1512.03385},
  year      = {2015},
  url       = {http://arxiv.org/abs/1512.03385},
  eprinttype = {arXiv},
  eprint    = {1512.03385},
  bibsource = {dblp computer science bibliography, https://dblp.org}
}

@article{hu2020weakly,
  title={Weakly supervised deep learning for covid-19 infection detection and classification from ct images},
  author={Hu, Shaoping and Gao, Yuan and Niu, Zhangming and Jiang, Yinghui and Li, Lao and Xiao, Xianglu and Wang, Minhao and Fang, Evandro Fei and Menpes-Smith, Wade and Xia, Jun and others},
  journal={IEEE Access},
  volume={8},
  pages={118869--118883},
  year={2020},
  publisher={IEEE}
}

@inproceedings{anwar2020deep,
  title={Deep learning based diagnosis of COVID-19 using chest CT-scan images},
  author={Anwar, Talha and Zakir, Seemab},
  booktitle={2020 IEEE 23rd International Multitopic Conference (INMIC)},
  pages={1--5},
  year={2020},
  organization={IEEE}
}

@inproceedings{tan2019efficientnet,
  title={Efficientnet: Rethinking model scaling for convolutional neural networks},
  author={Tan, Mingxing and Le, Quoc},
  booktitle={International conference on machine learning},
  pages={6105--6114},
  year={2019},
  organization={PMLR}
}

@article{yang2020covid,
  title={COVID-CT-dataset: a CT scan dataset about COVID-19},
  author={Yang, Xingyi and He, Xuehai and Zhao, Jinyu and Zhang, Yichen and Zhang, Shanghang and Xie, Pengtao},
  journal={arXiv preprint arXiv:2003.13865},
  year={2020}
}

@article{ahsan2020study,
  title={Study of different deep learning approach with explainable ai for screening patients with COVID-19 symptoms: Using ct scan and chest x-ray image dataset},
  author={Ahsan, Md Manjurul and Gupta, Kishor Datta and Islam, Mohammad Maminur and Sen, Sajib and Rahman, Md and Hossain, Mohammad Shakhawat and others},
  journal={arXiv preprint arXiv:2007.12525},
  year={2020}
}

@article{he2020sample,
  title={Sample-efficient deep learning for COVID-19 diagnosis based on CT scans},
  author={He, Xuehai and Yang, Xingyi and Zhang, Shanghang and Zhao, Jinyu and Zhang, Yichen and Xing, Eric and Xie, Pengtao},
  journal={medrxiv},
  year={2020},
  publisher={Cold Spring Harbor Laboratory Press}
}

@article{alshazly2021explainable,
  title={Explainable COVID-19 detection using chest CT scans and deep learning},
  author={Alshazly, Hammam and Linse, Christoph and Barth, Erhardt and Martinetz, Thomas},
  journal={Sensors},
  volume={21},
  number={2},
  pages={455},
  year={2021},
  publisher={Multidisciplinary Digital Publishing Institute}
}

@inproceedings{ye2021explainable,
  title={Explainable AI for COVID-19 CT classifiers: an initial comparison study},
  author={Ye, Qinghao and Xia, Jun and Yang, Guang},
  booktitle={2021 IEEE 34th International Symposium on Computer-Based Medical Systems (CBMS)},
  pages={521--526},
  year={2021},
  organization={IEEE}
}

@inproceedings{sundararajan2017axiomatic,
  title={Axiomatic attribution for deep networks},
  author={Sundararajan, Mukund and Taly, Ankur and Yan, Qiqi},
  booktitle={International conference on machine learning},
  pages={3319--3328},
  year={2017},
  organization={PMLR}
}

@inproceedings{selvaraju2017grad,
  title={Grad-cam: Visual explanations from deep networks via gradient-based localization},
  author={Selvaraju, Ramprasaath R and Cogswell, Michael and Das, Abhishek and Vedantam, Ramakrishna and Parikh, Devi and Batra, Dhruv},
  booktitle={Proceedings of the IEEE international conference on computer vision},
  pages={618--626},
  year={2017}
}

@inproceedings{cam,
  title={Learning deep features for discriminative localization},
  author={Zhou, Bolei and Khosla, Aditya and Lapedriza, Agata and Oliva, Aude and Torralba, Antonio},
  booktitle={Proceedings of the IEEE conference on computer vision and pattern recognition},
  pages={2921--2929},
  year={2016}
}

@inproceedings{ribeiro2016should,
  title={" Why should i trust you?" Explaining the predictions of any classifier},
  author={Ribeiro, Marco Tulio and Singh, Sameer and Guestrin, Carlos},
  booktitle={Proceedings of the 22nd ACM SIGKDD international conference on knowledge discovery and data mining},
  pages={1135--1144},
  year={2016}
}

@inproceedings{he2016deep,
  title={Deep residual learning for image recognition},
  author={He, Kaiming and Zhang, Xiangyu and Ren, Shaoqing and Sun, Jian},
  booktitle={Proceedings of the IEEE conference on computer vision and pattern recognition},
  pages={770--778},
  year={2016}
}

@inproceedings{huang2017densely,
  title={Densely connected convolutional networks},
  author={Huang, Gao and Liu, Zhuang and Van Der Maaten, Laurens and Weinberger, Kilian Q},
  booktitle={Proceedings of the IEEE conference on computer vision and pattern recognition},
  pages={4700--4708},
  year={2017}
}

@inproceedings{ronneberger2015u,
  title={U-net: Convolutional networks for biomedical image segmentation},
  author={Ronneberger, Olaf and Fischer, Philipp and Brox, Thomas},
  booktitle={International Conference on Medical image computing and computer-assisted intervention},
  pages={234--241},
  year={2015},
  organization={Springer}
}

@misc
{zenodo ,
title = {Zenodo},
url = {https://zenodo.org/record/3757476#.YqBY6KjMLIX},
journal = {},
author = {}
}

@misc
{medicalsegmentation ,
title = {Medical segmentation},
url = {http://medicalsegmentation.com/covid19/},
journal = {},
author = {}
}

@article{morozov2020mosmeddata,
  title={Mosmeddata: Chest ct scans with covid-19 related findings dataset},
  author={Morozov, Sergey P and Andreychenko, AE and Pavlov, NA and Vladzymyrskyy, AV and Ledikhova, NV and Gombolevskiy, VA and Blokhin, Ivan A and Gelezhe, PB and Gonchar, AV and Chernina, V Yu},
  journal={arXiv preprint arXiv:2005.06465},
  year={2020}
}

@article{khan2020covid,
  title={COVID-19 classification based on Chest X-Ray images using machine learning techniques},
  author={Khan, Naveed and Ullah, Farhat and Hassan, Muhammad Abul and Hussain, Adnan and others},
  journal={Journal of Computer Science and Technology Studies},
  volume={2},
  number={2},
  pages={01--11},
  year={2020}
}

@article{hosny2018artificial,
  title={Artificial intelligence in radiology},
  author={Hosny, Ahmed and Parmar, Chintan and Quackenbush, John and Schwartz, Lawrence H and Aerts, Hugo JWL},
  journal={Nature Reviews Cancer},
  volume={18},
  number={8},
  pages={500--510},
  year={2018},
  publisher={Nature Publishing Group}
}

@article{fang2020sensitivity,
  title={Sensitivity of chest CT for COVID-19: comparison to RT-PCR},
  author={Fang, Yicheng and Zhang, Huangqi and Xie, Jicheng and Lin, Minjie and Ying, Lingjun and Pang, Peipei and Ji, Wenbin},
  journal={Radiology},
  year={2020},
  publisher={Radiological Society of North America}
}

@misc
{keras ,
title = {keras},
url = {https://keras.io/api/applications/},
journal = {},
author = {}
}

@article{cohen2020covid,
  title={COVID-19 image data collection},
  author={Cohen, Joseph Paul and Morrison, Paul and Dao, Lan},
  journal={arXiv preprint arXiv:2003.11597},
  year={2020}
}

@article{haque2020deep,
  title={Deep learning approaches to biomedical image segmentation},
  author={Haque, Intisar Rizwan I and Neubert, Jeremiah},
  journal={Informatics in Medicine Unlocked},
  volume={18},
  pages={100297},
  year={2020},
  publisher={Elsevier}
}

@misc
{MedSeg ,
title = {MedSeg },
url = {https://htmlsegmentation.s3.eu-north-1.amazonaws.com/index.html},

}

@misc
{project,
title = {CT-xCOV},
url = {https://github.com/ismailelbouknify/CT-xCOV},

}

@incollection{el2021automated,
  title={Automated methods for detection and classification pneumonia based on x-ray images using deep learning},
  author={El Asnaoui, Khalid and Chawki, Youness and Idri, Ali},
  booktitle={Artificial intelligence and blockchain for future cybersecurity applications},
  pages={257--284},
  year={2021},
  publisher={Springer}
}

@article{hakkoum2022interpretability,
  title={Interpretability in the medical field: A systematic mapping and review study},
  author={Hakkoum, Hajar and Abnane, Ibtissam and Idri, Ali},
  journal={Applied Soft Computing},
  volume={117},
  pages={108391},
  year={2022},
  publisher={Elsevier}
}

@article{zhang2020clinically,
  title={Clinically applicable AI system for accurate diagnosis, quantitative measurements, and prognosis of COVID-19 pneumonia using computed tomography},
  author={Zhang, Kang and Liu, Xiaohong and Shen, Jun and Li, Zhihuan and Sang, Ye and Wu, Xingwang and Zha, Yunfei and Liang, Wenhua and Wang, Chengdi and Wang, Ke and others},
  journal={Cell},
  volume={181},
  number={6},
  pages={1423--1433},
  year={2020},
  publisher={Elsevier}
}

@inproceedings{fauvel2006combined,
  title={A combined support vector machines classification based on decision fusion},
  author={Fauvel, Mathieu and Chanussot, Jocelyn and Benediktsson, J},
  booktitle={2006 IEEE International Symposium on Geoscience and Remote Sensing},
  pages={2494--2497},
  year={2006},
  organization={IEEE}
}
\end{document}